\documentclass [a4paper,12pt]{article}
\usepackage{amssymb}
\usepackage{graphicx}
\usepackage{amsmath}
\usepackage{fancybox}
\usepackage{color}
\usepackage{geometry}

\newcommand{\be}{\begin{equation}}
\newcommand{\ee}{\end{equation}}
\newcommand{\bea}{\begin{eqnarray}}
\newcommand{\eea}{\end{eqnarray}} 
\newcommand{\rmd}{{\rm d}} 
\begin{document}
\setlength{\baselineskip}{18pt}
\begin{titlepage}

\begin{flushright}
%KOBE-TH-19-05   
NITEP 250
\end{flushright}
\vspace{0.3cm}
\begin{center}
{\LARGE\bf Dyons in higher-dimensional gauge theories} 
\end{center}
\vspace{5mm}

\begin{center}
{\large

Yuki Adachi, C.S. Lim$^a$\footnote{Professor Emeritus} and Nobuhito Maru$^{b, c}$
}
\end{center}

\vspace{0.3cm}

\centerline{{\it
Department of Sciences, Matsue College of Technology,
Matsue 690-8518, Japan.}}

\centerline{{\it
$^{a}$
Kobe University, Kobe 657-8501, Japan}} 

\centerline{{\it 
$^b$
Department of Physics, Osaka Metropolitan University,
Osaka 558-8585, Japan}}

\centerline{{\it 
$^c$
Nambu Yoichiro Institute of Theoretical and Experimental Physics (NITEP), }} 
\centerline{{\it Osaka Metropolitan University,  
Osaka 558-8585, Japan}}

\vspace{1.0cm}
\centerline{\large\bf Abstract}
\vspace{0.5cm}

We discuss the 't Hooft-Polyakov (TP) monopole and then dyon in the framework of higher dimensional gauge theories, 
such as gauge-Higgs unification models.
First, we point out that the Bogomol'nyi-Prasad-Sommerfield (BPS) monopole is nothing but a self-dual gauge field in the 4-dimensional (4D) space including the extra dimension, which is argued to lead to a consequence that the mass of the BPS monopole $M_{{\rm TPM}}$ and therefore the vacuum expectation value (VEV) of the Higgs field are topologically quantized. 
In literatures, there exist related arguments on the calorons, which may be understood to be a composition of a pair of constituent monopole and anti-monopole, with each constituent carrying fractional topological charge, while the net topological charge carried by the caloron is unity. From the viewpoint of the caloron, our conclusion of the quantized monopole mass corresponds to the special case, where only a single monopole exists that carries the net topological charge. 
Next, the argument is generalized to the case of dyon. The mass of the BPS dyon, $M_{{\rm BPS}}$, is still proportional to the quantized Higgs VEV, though it also depends on a parameter $\mu$, denoting the ratio of the electric and magnetic charges of the dyon. In the 5D gauge theories the Chern-Simons term is induced at the quantum level, which, after the extra space component of the gauge field is replaced by its VEV, produces the $\theta$ term. 
Then, through the Witten effect we reach to an interesting conclusion that the parameter $\mu$ and therefore $M_{{\rm BPS}}$ are discretized. 
In addition, we propose a numerical method to obtain the field configurations and the mass of the non-BPS dyons by use of ``modified" gradient flow equations.

\end{titlepage}

\section{Introduction} 

Solitons, especially the 't Hooft-Polyakov monopole \cite{'t Hooft, Polyakov} (TP monopole) has been playing important roles 
in the unified theories of particle physics. For instance, in the grand unified theory (GUT) based on a simple group SU(5), as one of 
the physics beyond the standard model (BSM), stable superheavy TP monopoles are overproduced early in the history of the Universe and 
contribute to the present energy density too generously. They will also induce catastrophic proton decays through the Rubakov effect \cite{Rubakov}. Among other things, to solve this problem by diluting the monopoles was one of the reasons why the inflationary cosmology was proposed.

As is well-known, the TP monopole can be naturally generalized in the same theoretical system with gauge field and Higgs field belonging to the adjoint representation of the gauge group (adjoint Higgs), so that the soliton carries not only a magnetic charge but also an electric charge \cite{Schwinger, Zwanziger, Julia-Zee}. This type of soliton is called dyon. 
Though the dyon is an interesting research object itself, it also comes on the stage almost inevitably, e.g., in the superstring theory or in the 4-dimentional (4D) N = 4 or N = 2 \cite{Seiberg-Witten} extended supersymmetric Yang-Mills (YM) theories, which may be 
regarded as the low-energy effective theories of the superstring theory. These theories possess an electric-magnetic duality: ``S-duality", $g \leftrightarrow \frac{4\pi}{g}$ ($g$: gauge coupling constant). 

On the other hand, higher dimensional theories with extra spatial dimension have been investigated from the viewpoint of unified theories, starting from the Kaluza-Klein (KK) theory which aimed to unify the gravity with electro-magnetism until the superstring theory which enables the unification of all four types of forces acting on the particles.              

In this paper, we focus on the ``gauge-Higgs unification" (GHU), which is one of the attractive scenarios of BSM theories. GHU is a theory formulated in the framework of higher dimensional gauge theories, where Higgs field is identified with the extra space component of the higher dimensional gauge field, to be more precise, its KK zero-mode \cite{Manton:1979, Hosotani}. 
In this way, gauge and Higgs interactions from the viewpoint of 4D space-time are naturally unified as higher dimensional gauge interaction. This is why it is called GHU.  
A remarkable attractive feature of GHU is that the gauge hierarchy problem, which is one of the crucial theoretical problem the standard model (SM) faces, is naturally resolved  \cite{HIL}, thanks to the higher dimensional local gauge symmetry, without relying on the supersymmetry (SUSY), thus opening a new avenue of BSM. 

In the 10D SUSY YM theory, i.e. the point particle limit of the (open) superstring, matter fields do not exist, and therefore Higgs field is inevitably identified with the extra space component of the gauge field. Namely, GHU also has an aspect as the effective low-energy theory of the superstring, though GHU does not necessitate SUSY, and its investigation has a potential to verify the specific properties of the superstring theory.

As we will discuss in this paper, the GHU turns out to be a natural framework to incorporate the TP monopole and dyon.  
For instance, the minimal unified electro-weak GHU model, incorporating the SM, is the 5D SU(3) model \cite{Kubo}. Since it has the simple group SU(3) as its gauge symmetry, there should exist the classical gauge field configuration of TP monopole, just as in the case of SU(5) GUT. 
It is also worth noting that in the GHU scenario the adjoint Higgs field, indispensable in order to construct the TP monopole, is automatically contained in the theory, as the extra space component of the gauge field, e.g. $A_y$ for 5D GHU, without any necessity of introducing additional scalar matter fields. 

By the way, in the GHU scenario, the vacuum expectation value (VEV) of the Higgs field is nothing but a constant gauge field with vanishing field strength (at least in 5D GHU), which seems to be a pure-gauge configuration having no physical meaning at the first glance. When, however, gauge fields exist only on a non-simply-connected space, constant gauge fields may be physically meaningful, as in the case of well-known Aharonov-Bohm (AB) effect. Similarly, for instance, in the simplified 5D GHU with a circle as its extra dimension, 
its Higgs field may be regarded as a sort of AB phase, or Wilson-line phase, more generally. As an example, in the case of 5D U(1) GHU model, the Wilson-loop along the circle is given as 
\be 
\label{1.1}
W = e^{ig \oint A_y \rmd y}, 
\ee 
where $y$ denotes the coordinate along the extra space and $A_y$, identified as the Higgs field, is the extra space component of the higher dimensional gauge field, with $g$ being the 5D gauge coupling constant. Here again, essential thing is that the circle has a topological property as a non-simply-connected space. Let us note that the Wilson-loop is a gauge invariant global operator, though it does not contain any differentials of the gauge field. This is why the Higgs field $A_y$ acquires its potential term $V(A_y)$ at quantum level as a function of $W$, although $V(A_y)$ is clearly forbidden by the local gauge invariance at the classical level. 
The fact that the Higgs filed behaves as a Wilson-line phase inevitably leads to a remarkable specific consequence of GHU (at least when the compact extra dimension is a non-simply-connected space such as a circle), which is never shared by the SM: all physical quantities are periodic in the Higgs field $A_y$. As an its typical example, the radiatively induced Higgs potential $V(A_y)$ is described by a 
trigonometric function of $A_y$, as we will see later.

It may be interesting to note that a similar periodicity is seen also in the potential of the pseudo Nambu-Goldstone bosons, 
such as the axion. Thus, the extra natural inflation scenario \cite{extra natural inflation, IKLM, Furuuchi, Hirose-Maru}, with $A_y$ being identified as the inflaton field, has some similarity to the natural inflation scenario \cite{natural inflation}, where the pseudo Nambu-Goldstone boson behaves as the inflaton, although in the case of the extra natural inflation the inflaton potential is not spoilt under the quantum gravity effect, since it is protected by the local gauge symmetry.

By the way, in the SM, in addition to the aforementioned gauge hierarchy problem, there exists another critical essential theoretical problem. Namely, it is not able to predict the observed masses of quarks and leptons and the generation mixing angles. Basic problem is that there is no guiding principle to constrain especially the values of Yukawa couplings relevant for the fermion masses and mixings as seen in the Higgs interaction. 
The GHU scenario has a potential to solve this problem too, since the Higgs interactions are constrained by gauge principle. On the other hand, it seems to be quite non-trivial to realize, e.g., the observed impressive hierarchical structure of the fermion masses depending on their generation, just because the gauge interaction is governed by a single universal parameter, i.e. gauge coupling constant.

As a first step to resolve this problem, in \cite{mass hierarchy} a 6D GHU model was considered. Placing a monopole at a point inside the torus T$^2$, i.e. its extra dimension, the hierarchical fermion masses have been demonstrated to be realized naturally, 
without any tuning of parameters. Here, the hierarchical fermion masses mean that when the logarithm of the masses for each of fermion species, e.g. charged lepton or up-type quark, are plotted as a function of generation number, the plotted points align approximately along a line. 
In addition, three generations of fermion emerge as KK zero-modes, which is the consequence of the quantization of the magnetic charge, as implied by the index theorem. In \cite{mass hierarchy}, however, the monopole was assumed to be of the Dirac-type, having an infinite mass. Clearly, it will be desirable if it can be replaced by the TP monopole, without any singularity in the gauge field configuration and therefore having a finite mass, which naturally exists in the GHU scenario.   

Being motivated by these viewpoints mentioned above, in this paper, we reconsider the TP monopole and then dyon in the theoretical framework of the higher dimensional gauge theories, such as GHU. As has been already stated, in these theories, the adjoint Higgs 
field, indispensable for the construction of TP monopole or dyon, has been built in as the extra space component of the gauge field, 
and its interaction is strongly constrained by the gauge principle. Furthermore, as a specific feature of this type of theories, 
the Higgs potential is periodic in the Higgs field. This fact is quite different from what we expect in the ordinary 4D space-time and 
plays a crucial role in the characteristic predictions made by the higher dimensional gauge theories, as discussed later.

First, we discuss that, as a specific feature of higher dimensional gauge theory, the Bogomol'nyi-Prasad-Sommerfield (BPS) 
saturated TP monopole (say, BPS monopole), which 
satisfies the BPS condition and therefore the equation of motion justified for the case of negligible Higgs potential, is nothing but an (anti-)self-dual gauge field in the 4D space with coordinates $(\vec{x}, y)$ including the extra space (not 4D space-time) \cite{Hasegawa-Lim}. 
This is not just the matter of interpretation, but it is demonstrated to have an interesting consequence. 
Namely, just as the action of the instanton, an (anti-)self-dual gauge field in the ordinary 4D space-time, is quantized ($\pi_{3}(SU(2)) = Z$), the mass of the BPS monopole $M_{{\rm TPM}}$ (TPM stands for the TP monopole) is known to be quantized, being described just by the higher dimensional gauge coupling $g$ as $M_{{\rm TPM}} = \frac{8\pi^{2}}{g^{2}}$, corresponding to the Pontryagin index $\nu = -1$ (see eq.(\ref{2.9})). Or, in other words, the VEV of the Higgs field, or equivalently that of $A_y$, is quantized to a specific value from the 4D point of view. This is in clear contrast to the case of ordinary 4D space-time where $M_{{\rm TPM}}$ is proportional to the Higgs VEV, which is left as a free parameter, since the Higgs potential to fix the VEV is absent for the BPS monopole.

In literatures, there are related discussions concerning calorons, which are instanton-like self-dual 
gauge fields, that are periodic in the compactified Euclidean time \cite{Harrington, Lee, Dunne}. Interestingly, it has been shown that a caloron can be understood to be a composition of a pair of constituent monopole and anti-monopole , where each constituent carries fractional topological charge, while the net topological charge carried by the caloron itself is unity \cite{Lee}. 
As is discussed in more details in the next section, from the viewpoint of the caloron, our conclusion of the quantized monopole mass corresponds to the specific case, where only a single monopole exists that carries the net integer topological charge.     

Next, we generalize such argument to the case of dyon. It turns out that when the Higgs potential is absent, one of the BPS conditions still can be written as an (anti-)self-dual-like condition, similar to that in the case of BPS monopole. 
The analytic dyon solution to satisfy these conditions (say, BPS dyon) is shown to be easily constructed by use of the functions utilized to describe the BPS monopole. 
The mass of the dyon is still proportional to the quantized Higgs VEV in the context of the higher dimensional gauge theory. 
It, however, now also depends on a parameter $\mu$, denoting the ratio of the electric and magnetic charges of the dyon 
($\tan \mu = \frac{q}{g_{m}}$), which is usually left as a free parameter.  

In this paper, we propose a mechanism to fix $\mu$, which again utilizes characteristic features of higher dimensional 
gauge theories. 
Let us note that dyons, possessing both of electric and magnetic charges, may be understood to be P and CP-violating field configurations. Witten has proposed an interesting way to fix the electric charge $q$ of dyons in the theory 
containing the $\theta$ term, $\theta F_{\mu \nu}\tilde{F}^{\mu \nu}$, which clearly violates both of P and CP symmetries: 
it has been shown that $q$ is fixed in proportion to $\theta$ \cite{Witten}.  
In \cite{Witten}, however, the origin of the $\theta$ term was not specified. 
We propose an approach to attribute the origin to the radiatively induced Chern-Simons (CS) term of the form 
$C_y F_{\mu \nu}\tilde{F}^{\mu \nu}$ in a 5D gauge theory, where $C_y$ denotes the extra space component of additional 
U(1) gauge field, independent of $A_\mu, \ A_y$ to form the dyon solution. Then the VEV of $C_y$ clearly behaves as $\theta$. 
A characteristic feature of this approach is that the VEV of $C_y$ is quantized, just as in the case of the VEV of $A_y$. 
As the result, $\theta$ and therefore $q$ and finally $\mu$ are quantized through the ``Witten effect", a feature not shared 
by ordinary 4D theories. 
Thus we reach to an interesting conclusion that the mass of the dyon only takes discrete values.    
    
We will also discuss the dyon solution and its mass when the Higgs potential is present (say, non-BPS dyon). 
In the GHU scenario, though the Higgs potential is forbidden at the classical lagrangian by the local gauge invariance, it is induced at the quantum level as a function of the gauge invariant global operator, i.e. Wilson-loop, thus leading to the periodicity of the potential, as has been already stated above. Unfortunately, the analytic solution of the equation of motion is not known for the non-BPS dyon. Hence, in order to get the gauge field configuration and the mass of the non-BPS dyon, some numerical method is necessitated.

In this paper, we adopt the method of gradient flow to obtain the field configuration and the mass of the dyon numerically. 
For instance, in the simplified case of the BPS monopole (i.e. $\mu = 0$), it is shown that by utilizing the gradient of the Hamiltonian in the gradient flow equation, we can readily reconstruct the analytic solutions correctly. 
In the case of the dyon ($\mu \neq 0$), however, because of the contribution of the ``electric" field (though the gauge group is non-Abelian), the extremum conditions of lagrangian $L$, or rather $-L$, and the Hamiltonian $H$ do not coincide, and the gradient of $- L$ (not $H$) should be adopted to get the solution of the equation of motion correctly. 
However, such approach turns out to face a problem: it causes an unstable behavior in the flow of the field configuration. The reason for the instability should be the fact that the contribution of the ``electric" field to $- L$ is semi-negative-definite. 

To overcome this problem, in this paper we propose a modified gradient flow equations, where concerning the equation of motion for $A_0$, which is responsible for the instability, ``Gauss law constraint", we prepare a separated gradient flow equation. Such modified ``two-step gradient flow equations" (see the text below for the details) are confirmed to produce satisfactory field configurations and the mass of non-BPS dyon, as we will see below. 
As the result we find, in particular, that the deviation of the field configurations and the mass from the case of BPS dyon is relatively small, probably reflecting the fact that the periodic Higgs potential is induced at the quantum level.

%%%%%%%%%%%%%%%%%%%%%%%%%%%%%%%%%%%%%%%%%%%%%%%%%
\section{The BPS monopole as a self-dual field in 5D gauge theory} 
%%%%%%%%%%%%%%%%%%%%%%%%%%%%%%%%%%%%%%%%%%%%%%%%%
%%%%%%%%%%%%%%%%%%%%%%%%%%%%%%%%%%%%%%%%%%%%%%%%% 

Before discussing dyons, we first argue that the BPS saturated TP monopole, say BPS monopole, 
is able to be understood as an (anti-)self-dual gauge field in 5D SU(2) GHU \cite{Hasegawa-Lim}.\footnote{Recently, 
cosmic string, which is another solution of the soliton, has been studied in a similar setup \cite{HiroseKanda}.} 
It should be noted that in the 5D theory, the adjoint Higgs field $\phi$, indispensable for the construction of TP monopole has been built in as the extra space component of the 5D gauge field, $A_{y}$.

The (classical) 5D action to start with is just that of the 5D SU(2) pure Yang-Mills gauge theory: 
\be 
\label{2.1} 
S =\int {\cal L} \ \rmd^{4}x \rmd y, \ \ \ \ \ {\cal L} = - \frac{1}{2}{\rm
Tr}\left(F_{MN}F^{MN}\right) \ \ \ \ \ \left(M, N = \mu, y \ (\mu = 0-3)\right),  
\ee 
where $A_{M} = A^{(a)}_{M} \frac{\tau_{a}}{2}$ and $F_{MN} = \partial_{M}A_{N} - \partial_{N}A_{M} -ig [A_{M}, A_{N}]$. 
The compactified extra dimension with coordinate $y$ is assumed to be a circle of radius $R$.

In the ordinary 4D space-time, the BPS monopole is well-known to be an analytic static solution of the following BPS condition:  
\be 
\label{2.2} 
F_{ij} = - \epsilon_{ijk} D_{k}\phi \ \ (i,j,k = 1-3),     
\ee 
where $D_{k}\phi \equiv \partial_{k} \phi -ig [A_{k}, \phi]$.  
This is also the solution to the equation of motion, derived from (\ref{2.1}) without the potential of $\phi$.     

Interestingly, once we identify $A_{y}$ with $\phi$, as is done in the GHU scenario, the BPS condition turns out to be nothing but an (anti-)self-dual condition for the gauge fields in the 4D space with spatial coordinates $(x_{i}, y) \ (i = 1-3)$ (instead of the ordinary 4D space-time) \cite{Hasegawa-Lim}:  
\be 
\label{2.3} 
F_{IJ} = - \frac{1}{2}\epsilon_{IJKL} F_{KL} \ \ (I,J,K,L = i,y ),    
\ee 
with $\epsilon_{123y} = 1$. (A similar observation was made in the literature \cite{Manton}, but in the framework of the gauge theory in the ordinary 4D space-time, not in the framework of GHU).
In fact, if we retain only the KK zero modes of 5D gauge fields, neglecting their $y$-dependence \footnote{To be precise, the KK non-zero modes of $A_y$ are unphysical, in the sense that they are absorbed by the corresponding KK non-zero modes of $A_{\mu}$ as their longitudinal components, just as in the Higgs mechanism, and only the KK zero mode of $A_{y}$ may be regarded as physical.}, since we are interested in the low-energy effective theory of (\ref{2.1}), then $F_{ky} = D_{k}A_{y} \equiv \partial_{k} A_{y} -ig [A_{k}, A_{y}] \ (\partial_y A_{k} = 0)$, which may be interpreted as $D_{k}\phi$.
The self-duality (\ref{2.3}) also may be written as 
\be 
\label{2.4} 
B_{k} = - F_{ky},      
\ee 
where $B_{k} \equiv \frac{1}{2}\epsilon_{kij} F_{ij}$ is a ``magnetic" field, though it is not a field strength of Abelian gauge field.

To be precise, the self-duality condition should be accompanied by the following subsidiary constraints, 
\be 
\label{2.5}
F_{0I} = 0: \ \ \ \ F_{0i} \equiv E_{i} = 0, \ \ F_{0y} = 0,
\ee 
where $E_{i}$ is an ``electric" field. These constraints are easily satisfied for static ($t$-independent) KK zero modes of the fields, also assuming $A_{0} = 0$.    

In the case of the BPS monopole in the framework of ordinary 4D gauge theory, its mass denoted by $M_{{\rm TPM}}$ is given by a familiar formula 
\be 
\label{2.6}  
M_{{\rm TPM}} = g_{m} v_{4} = \frac{4\pi v_{4}}{g_{4}}, 
\ee 
where 
\be 
\label{2.6a}
g_{m} = \frac{4\pi}{g_{4}}
\ee 
is the well-known quantized magnetic charge and $g_{4}$, $v_{4}$ are the 4D gauge coupling and the magnitude of the VEV of $\phi$, related to the 5D counterparts $g, \ v$ by multiplying suitable powers of the factor $\sqrt{2\pi R}$. In the  ordinary 4D space-time, there is no way to fix the mass $M_{{\rm TPM}}$, since $v_{4}$ is left as a free parameter. Note that the potential of $\phi$, to fix the VEV, is assumed to be absent for the case of BPS monopole.   

In the framework of 5D GHU, however, since the BPS monopole is interpreted as a self-dual gauge field, we naturally expect that its mass cannot be arbitrary but is topologically quantized, just as the action of the instanton is quantized in 4D space-time. 

We now demonstrate that it really is the case. Under the subsidiary constraints (\ref{2.5}), the Hamiltonian of the theory          
\be 
\label{2.7}
H = \int {\cal H} \ \rmd^{3}x \rmd y, \ \ \ \ \ {\cal H} = \frac{1}{2}{\rm
Tr}\left(F_{IJ}^{2}\right) 
= \frac{1}{4}{\rm Tr}\left[ \left(F_{IJ} + \tilde{F}_{IJ}\right)^{2} - 2
F_{IJ}\tilde{F}_{IJ} \right],  
\ee 
where $\tilde{F}_{IJ} \equiv \frac{1}{2} \epsilon_{IJKL} F_{KL}$. In the r.h.s. of the last equality, 
the first term just vanishes because of the (anti-)self-duality (\ref{2.3}) and the remaining term is well-known to be topologically quantized, just as the action of the instanton solution is quantized in ordinary 4D space-time:   
\be 
\label{2.8}
\frac{g^{2}}{16\pi^{2}} \int {\rm Tr}\left(F_{IJ}\tilde{F}_{IJ}\right) \ \rmd^{3}x \rmd y = \nu \ \ \ (\nu: {\rm integer}),   
\ee 
where $\nu$ is a winding number, known as Pontryagin index. 
Note that the combination $gA_{I}$ is invariant under the change of the space-time dimensionality and we can freely commute between 4D and 5D space-time. So the well-known relation, similar to (\ref{2.8}), in ordinary 4D space-time can be readily 
applied for our 5D gauge theory as well.    

We thus get a quantized energy and $\nu = - 1$ leads to the mass of the BPS monopole 
\be 
\label{2.9}
M_{{\rm TPM}} = \frac{8\pi^{2}}{g^{2}} = \frac{4\pi}{g_{4}^{2}} \frac{1}{R}, 
\ee 
where the 5D gauge coupling $g$ is not a dimensionless quantity and can be replaced by the dimensionless 4D coupling through a relation $g_{4} = g/\sqrt{2\pi R}$. 
In this way, $M_{{\rm TPM}}$ is completely fixed, being topologically quantized. 
In other words, $v_{4}$ appearing in (\ref{2.6}) is also quantized, without referring to the Higgs potential, in the framework of 5D GHU.

In literatures, there exist arguments on the related topics, i.e. calorons, which are instanton-like (anti-)self-dual 
gauge fields, that are periodic in the compactified Euclidean time, with a period $\beta = \frac{1}{T}$ denoting the inverse temperature in the context of finite temperature gauge theory \cite{Harrington, Lee, Dunne}. 
Though calorons are periodic instantons rather than monopoles, interestingly, in ref.\cite{Lee} it has 
been shown that a caloron may be understood to be a composition of a pair of constituent monopole and anti-monopole with opposite magnetic charges. In the argument, each constituent is regarded to carry 
fractional topological charge $k_{1}$ or $k_2$ (corresponding to the Pontryagin index) handled by $u$, the VEV of $A_4$ (corresponding to $gA_y$ in our case), 
\be 
\label{2.9a} 
k_{1} = \frac{\beta u}{2\pi}, \ \ k_{2} = 1 - \frac{\beta u}{2\pi}, 
\ee 
though the net topological charge carried by the caloron is unity, $k_{1}+k_{2} = 1$. 
\newpage

The difference of our argument from the argument for the caloron given above is that in our case the self-dual field is for a single monopole, not for a pair of monopoles, which itself carries the 
integer topological charge $|\nu| = 1$. 
From the viewpoint of the caloron, the situation in our paper corresponds to the specific case $u = \frac{2\pi}{\beta}$, i.e. $k_1 = 1, \ k_{2} = 0$. As is discussed below, in the GHU scenario, the VEV of $A_y$ cannot be arbitrary but is shown to be discretized relying on a general argument.

Since this BPS monopole is the minimum energy configuration for a fixed $\nu$, it should be the solution of the equation of motion for the gauge fields. Or, we may also say that the self-dual field automatically satisfies the equation of motion (derived from (\ref{2.1})) by virtue of the Bianchi identity.

As a matter of fact, this quantization of $M_{{\rm TPM}}$ can be endorsed by an argument from a different point of view, i.e. on the Higgs VEV. 
Since the VEV of $\phi$ in (\ref{2.6}) is actually given by the corresponding VEV $\langle A_y \rangle$ in our 5D theory, it should satisfy 
\be 
\label{2.10}
g |\langle A_y \rangle| = \frac{1}{R}, 
\ee 
for (\ref{2.6}) and (\ref{2.9}) to be consistent with each other.

Although the classical action (\ref{2.1}) does not 
contain the potential term for $A_y$, $V(A_y)$, being forbidden by the local gauge invariance, it is known to be generated at the quantum level. 
We can confirm that the VEV given in (\ref{2.10}) is precisely what we get by the minimization of $V(A_y)$. 
The essence of the argument here is that the VEV is completely fixed,
irrespectively of the detail of the theory. Let us note that for SU(2) doublet
matter field, the Wilson-loop ${\rm exp}(i g \oint A_{y} \rmd y) \ (A_{y} = A^{(a)}_{y}\frac{\tau_{a}}{2})$ 
has a periodicity with respect to $A^{(a)}_{y}$ with a period $\frac{2}{gR}$. Hence $V(A_y)$, which is regarded as the function of the Wilson-loop, also has such periodicity: 
\be 
\label{2.11}
V\left(A^{(a)}_y + \frac{2}{gR}\right) = V\left(A^{(a)}_y\right).
\ee 
On the other hand, the potential is expected to be an even function of $A^{(a)}_y$, since $A^{(a)}_{y}$, which couples with fermion together with $\gamma_{5}$, may be regarded as a pseudo-scalar from 4D viewpoint, while the theory itself, adopting the compactification on the circle, is expected to preserve the parity symmetry: 
\be 
\label{2.12}  
V\left(-A^{(a)}_{y}\right) = V\left(A^{(a)}_{y}\right). 
\ee  
Then, combining (\ref{2.11}) and (\ref{2.12}), it is easily seen that the potential is even under the reflection with respect to the point $A^{(a)}_{y} = \frac{1}{gR}$: 
\be 
\label{2.13}
V\left(\frac{1}{gR} + X\right) = V\left(-\frac{1}{gR} - X\right) =
V\left(\frac{1}{gR} - X\right) \ \ \ \ \ (X \equiv A^{(a)}_{y} - \frac{1}{gR}), 
\ee 
which means that $A^{(a)}_{y} = \frac{1}{gR}$ is at least an extremum of the potential, implying $g |\langle A_y \rangle| = \frac{1}{R}$, i.e. (\ref{2.10}). (If $A^{(a)}_{y} = \frac{1}{gR}$ provides the maximum of the potential, then another extremum $A^{(a)}_{y} = 0$ will provide the minimum, but such a case is not of our interest.) 
It is interesting to note that the periodicity of the potential is again the reflection of the topological property of the extra dimension, i.e. the circle is non-simply-connected space and $V(A_y)$ can be regarded as the function of a sort of AB phase (Wilson-line phase).       

In fact, the 5D potential induced by the quantum correction due to the exchange of a doublet fermion, such as $\psi$ introduced later, with 
a bulk mass $M$, is calculated to be \cite{Kubo, Maru-Takenaga}
\be 
\label{2.14}
V(A_y) = \frac{6}{\pi^{2}L^{5}} \sum_{w = 1}^{\infty} \frac{1}{w^{5}}\left[ 1 +
wML + \frac{1}{3}(wML)^{2} \right] e^{- wML} \cos \left(w g\pi
RA^{(3)}_{y}\right), 
\ee 
where $L \equiv 2\pi R$ and $w$ denotes the winding number (not KK mode) appearing in the procedure of Poisson re-summation, and the third component of the triplet gauge field $A^{(3)}_y$ has been chosen to have the VEV, without loss of generality.
(\ref{2.14}) clearly has the properties of (\ref{2.11}) and (\ref{2.12}).
Now we confirm explicitly that the potential really leads to $g |\langle A_y \rangle| = g|\langle A^{(3)}_y \rangle| = \frac{1}{R}$.

%%%%%%%%%%%%%%%%%%%%%%%%%%%%%%%%%%%%%%%%%%%%%%%%%
\section{Dyon as a self-dual-like field in 5D gauge theory} 
%%%%%%%%%%%%%%%%%%%%%%%%%%%%%%%%%%%%%%%%%%%%%%%%%
%%%%%%%%%%%%%%%%%%%%%%%%%%%%%%%%%%%%%%%%%%%%%%%%% 

We are now ready to discuss dyon \cite{Julia-Zee}, which is the generalization of TP monopole so that it also carries an electric charge, in the framework of the 5D SU(2) GHU.

Among the subsidiary constraints (\ref{2.5}), we still retain 
\be 
\label{3.1} 
F_{0y} = 0,  
\ee 
while we cannot impose the remaining constraint $F_{0i} = E_{i} = 0$ (nor $A_{0} = 0$), as the dyon is supposed to be electrically charged. 
We instead impose the following ``Gauss law constraint", which is nothing but the equation of motion concerning $A_0$, under (\ref{3.1}): 
\be 
\label{3.2} 
D_{I}F_{0I} = 0: \ \ \ D_{i}E_{i} = 0. 
\ee 

Under the constraint (\ref{3.1}), the Hamiltonian of the theory is written in terms of ``electric" and ``magnetic" fields $E_{i}$ and $B_{i}$ as          
\be 
\label{3.3}
H = \int \ {\rm Tr} \left( E^{2}_{i} + B^{2}_{i} + F^{2}_{iy} \right) \
\rmd^{3}x \rmd y.  
\ee 
In this section, we ignore the potential $V(A_y)$, and consider ``BPS dyon". 
Since $H$ is symmetric under the dual transformation $E_{i} \leftrightarrow B_{i}$, we are tempted to 
move to the base of $E'_{i}, \ B'_{i}$ defined by the following SO(2) rotation by an angle $\mu$, which turns out to denote 
the ratio of the electric charge $q$ to the magnetic charge $g_m$ of the dyon \cite{Manton}, 
\be 
\label{3.4}
\begin{pmatrix}  
E'_{i} \\ 
B'_{i} 
\end{pmatrix} 
= 
\begin{pmatrix} 
\cos \mu  & -\sin \mu \\ 
\sin \mu & \cos \mu 
\end{pmatrix} 
\begin{pmatrix}  
E_{i} \\ 
B_{i} 
\end{pmatrix}.    
\ee 
Then, we rewrite the Hamiltonian, by a manipulation similar to (\ref{2.7}) in the case of BPS monopole:  
\bea 
H &=& \int \ {\rm Tr} \left( E^{2}_{i} + B^{2}_{i} + F^{2}_{iy} \right) \
\rmd^{3}x \rmd y \nonumber \\ 
 &=& \int \ {\rm Tr} \left( E'^{2}_{i} + B'^{2}_{i} + F^{2}_{iy} \right) \
 \rmd^{3}x \rmd y \nonumber \\ 
 &=& \int \ {\rm Tr} \left[ E'^{2}_{i} + (B'_{i} + F_{iy})^{2} - 2
 B'_{i}F_{iy} \right] \
 \rmd^{3}x \rmd y.  
\label{3.5} 
\eea 

Now we focus on the last term of the right hand side (r.h.s.) of (\ref{3.5}): 
\be 
\label{3.6} 
2 \int \ {\rm Tr} \left(- B'_{i}F_{iy}\right) \ \rmd^{3}x \rmd y 
= -2 \int \ {\rm Tr} \left[(
\cos \mu B_{i} + \sin \mu E_{i})F_{iy} \right] \ \rmd^{3}x \rmd y. 
\ee 
The first term of the r.h.s. can be expressed in terms of the magnetic charge $g_{m}$, since by use of the trivial relation $D_{i}B_{i} = 0$ (Bianchi identity),  
\be 
\label{3.7} 
-2 \int \ {\rm Tr} \left(B_{i}F_{iy}\right) \ \rmd^{3}x \rmd y = -2 \int \ {\rm
Tr} \left(B_{i}D_{i}A_{y}\right) \ \rmd^{3}x \rmd y
= -2 \int \ {\rm Tr} \partial_{i}\left(B_{i}A_{y}\right)  \rmd^{3}x \rmd y = v_{4}g_{m}.  
\ee 
Let us note that the space-volume integral is equivalent to the surface integral of $B_{i}A_{y} = v \cdot (\frac{A_{y}}{v}B_{i})$ at spatial infinity, where $\frac{A_{y}}{v}B_{i}$ is the magnetic field corresponding the U(1) symmetry, which remains after the spontaneous symmetry breaking due to the hedgehog-type field configuration of $A_y$, and its surface integral should correspond to the magnetic charge, though actually $\frac{A_{y}}{v}B_{i}$ is negative quantity corresponding to the negative winding number $\nu = -1$ in (\ref{2.8}). 

From 5D point of view, (\ref{3.7}) is easily understood by a topological argument: since the topological property does not 
change even if the fields $A_{i}, \ A_{y}$ themselves in the case of dyon are modified from the case of BPS monopole,  
(\ref{2.8}) still tells us  
\be 
- 2 \ \int \ \ {\rm Tr} \left( B_{i}F_{iy} \right) \ \rmd^{3}x \rmd y = - \frac{8\pi^{2}}{g^{2}} \nu.    
\label{3.8} 
\ee 
Setting $\nu = - 1$, and using $g^{2} = (2\pi R)g_{4}^{2}, \ v_{4} = \frac{1}{g_{4}R}$ and $g_{m} = \frac{4\pi}{g_{4}}$ (as is explicitly shown in (\ref{3.23}), the magnetic charge of the dyon does not change from the one given by (\ref{2.6a}) for the case of TP monopole), this just coincides with (\ref{3.7}).

Concerning the second term of the r.h.s. of (\ref{3.6}), thanks to the Gauss law constraint $D_{i}E_{i} = 0$ (\ref{3.2}), we can repeat a similar argument and it can be written in terms of the electric charge $q$ of the dyon:   
\be 
\label{3.9} 
-2 \int \ {\rm Tr} \left(E_{i}F_{iy}\right) \ \rmd^{3}x \rmd y = v_{4}q.   
\ee 
Then, from (\ref{3.5}) and (\ref{3.6}), we obtain an inequality for the energy, 
\be 
\label{3.10} 
E \geq v_{4} (\cos \mu \ g_{m} + \sin \mu \ q). 
\ee 
The equality is realized for field configurations satisfying the following relations: 
\bea 
&& E'_{i} = 0: \ \ \ \cos \mu \ E_{i} = \sin \mu \ B_{i},  
\label{3.11} \\ 
&& B'_{i} = - F_{iy}: \ \ \cos \mu \ B_{i} + \sin \mu \ E_{i} = - F_{iy}.  
\label{3.12}
\eea 
Th relation (\ref{3.12}) is an ``(anti-)self-dual-like" relation, though, containing a piece of $E_{i}$, it is not an 
exact self-duality except for $\mu = 0$.  

From (\ref{3.11}), we immediately conclude that the angle $\mu$ actually corresponds to the ratio of $g_{m}$ and $q$, as 
\be 
\label{3.13} 
\cos \mu \ q = \sin \mu \ g_{m} \ \ \to \ \ \tan \mu = \frac{q}{g_{m}}.  
\ee 
Thus the minimum of the energy for given $g_{m}$ and $q$, namely the mass of the BPS dyon, denoted by $M_{{\rm BPS}}$, is found to be 
\bea 
M_{{\rm BPS}} &=& v_{4} (\cos \mu \ g_{m} + \sin \mu \ q) = \frac{1}{\cos \mu}v_{4}g_{m} 
= \frac{1}{\cos \mu} M_{{\rm TPM}} \nonumber \\ 
&=& \frac{1}{\cos \mu} \frac{8\pi^{2}}{g^{2}} = \frac{1}{\cos \mu}\frac{4\pi}{g_{4}^{2}} \frac{1}{R}. 
\label{3.14} 
\eea 

The field configurations satisfying (\ref{3.11}) and ({\ref{3.12}}) can be constructed rather easily by use of the functions to describe field configurations for the case of BPS monopole \cite{Manton}, as we demonstrate now.
First, combining (\ref{3.11}) with (\ref{3.12}), we get 
\be 
\label{3.15}
B_{i} = - \cos \mu F_{iy}.
\ee 
Since $F_{iy} = D_{i} A_{y}$ is linear in $A_{y}$, we readily obtain a solution for $A_{y}$ to this condition, by dividing the field configuration of $A_{y}$ in the case of BPS monopole by $\cos \mu$, while retaining $A_{i}$. 
However, then $|A_y|$ behaves as $\frac{v}{\cos \mu}$, not $v$, at spatial infinity. This problem is easily resolved by replacing $v$  by $v \cos \mu$, since 
the solution of BPS monopole holds for arbitrary $v$, as we pointed out earlier. 
Namely, the solution to (\ref{3.15}) is given by  
\bea 
A_y &=& F(\rho) v \hat{x}_{a}\frac{\tau_{a}}{2}, \\ 
gA_{i} &=& G(\rho) \frac{1}{r} \epsilon_{aij} \hat{x}_{j}\frac{\tau_{a}}{2},
\label{3.16}
\eea 
where $\rho \equiv gvr \cos \mu$, $\hat{x}_{i} = \frac{x_{i}}{r} \ \left(r =
\sqrt{x^{2}_{i}}\right)$ and  
\bea 
F(\rho) &=&  \coth \rho - \frac{1}{\rho},  
\label{3.17}   \\ 
G(\rho) &=& 1 - \frac{\rho}{\sinh \rho}.  
\label{3.18}
\eea 

Next, combining (\ref{3.11}) with (\ref{3.12}) we also get 
\be 
\label{3.19}
E_{i} = - \sin \mu F_{iy}.  
\ee  
Let us note that both of the static KK zero modes of $A_0$ and $A_y$ behave as if they were just adjoint 
matter fields: $E_{i} = - D_{i}A_{0}, \ F_{iy} = D_{i}A_{y}$. Therefore (\ref{3.19}) readily holds for $A_{0}$, 
satisfying   
\be 
\label{3.20}
A_{0} = \sin \mu \ A_{y}.   
\ee 
Namely, writing $A_{0}$ in a form similar to $A_{y}$, since both fields behave as adjoint 
matter fields,  
\be 
\label{3.21}
A_{0} =  J(\rho) v \hat{x}_{a}\frac{\tau_{a}}{2},   
\ee
the function $J(\rho)$ is given in terms of $F(\rho)$ as   
\be 
\label{3.22} 
J(\rho) =  \sin \mu \ F(\rho) =  \sin \mu \ \left( \coth \rho - \frac{1}{\rho} \right). 
\ee 

Now the remaining constraint (\ref{3.1}), i.e. $F_{0y} = 0$, is automatically satisfied since fields are static KK zero modes 
and also $[A_{0}, A_{y}] = 0$ under (\ref{3.20}). 
Thus we have obtained the field configuration of BPS dyon, utilizing the BPS monopole solution, modified by the angle $\mu$, which is so far left as an arbitrary parameter.

Though the field configuration of BPS dyon is not exactly self-dual for $\mu \neq 0$, it is still rather easy to show, without performing explicit calculations, that it satisfies the equations of motion. For instance, the equation of motion for $A_y$, i.e. $D_{i}F_{iy} = 0$, is known to be satisfied, thanks to the ``self-dual-like" relation (\ref{3.15}), $B_{i} = - \cos \mu F_{iy}$, and the Bianchi identity: $D_{i}F_{iy} = - \frac{1}{\cos \mu}D_{i}B_{i} = 0$.

One remark here is that, as we have already used the fact, since the magnetic charge $g_{m}$ is topologically quantized, we should have the same quantized value $g_{m} = \frac{4\pi}{g_{4}}$ as in the case of TP monopole, although the argument $\rho$ of $F(\rho), \ G(\rho)$ has been rescaled by the factor $\cos \mu$. 
This can be explicitly shown by calculating a surface integral of the magnetic flux $- \frac{1}{v} 2{\rm Tr} (A_{y}B_{i})$, written in terms of $F, G$, on the sphere ${\rm S}^{2}$ located at spatial infinity $r \to \infty$: 
\be 
\label{3.23} 
g_{m} = \int_{{\rm S}^{2}} F(\rho)\frac{2G(\rho) - G(\rho)^{2}}{g_{4}r^{2}} \rmd S =  \frac{1}{g_{4}r^{2}}\cdot 4\pi r^{2} =  \frac{4\pi}{g_{4}},  
\ee 
where $\lim_{r \to \infty}F(\rho) = \lim_{r \to \infty}G(\rho) = 1$ have been used.

%%%%%%%%%%%%%%%%%%%%%%%%%%%%%%%%%%%%%%%%%%%%%%%%%
\section{Induced Chern-Simons term and quantized electric charge of dyon}   
%%%%%%%%%%%%%%%%%%%%%%%%%%%%%%%%%%%%%%%%%%%%%%%%%
%%%%%%%%%%%%%%%%%%%%%%%%%%%%%%%%%%%%%%%%%%%%%%%%% 

As we have seen in the previous two sections, in the framework of the 5D gauge theory the mass of the BPS monopole is completely fixed, being topologically quantized. We also have learnt that similar argumentation holds for the case of dyon and the dyon mass is 
topologically fixed up to the freedom of newly introduced angle $\mu$, the parameter to denote the ratio of electric and magnetic charges of the dyon: $\tan \mu = \frac{q}{g_{m}}$.

Witten has proposed an interesting elegant mechanism to fix the electric charge $q$ \cite{Witten}. Let us note that the coexistence of electric and magnetic charges means some violation of P and CP symmetries. Witten introduced the $\theta$ term into the original theory:
\be 
\label{4.1}
{\cal L}_{\theta} = \frac{g^{2}}{16\pi^{2}} \theta \ {\rm Tr} \left(F_{\mu \nu}\tilde{F}^{\mu \nu}\right),   
\ee 
which clearly breaks both symmetries. 
Note that this term may be written as a total derivative and does not affect the equation of motion.         

Then, he constructed an operator, corresponding to the generator of U(1) or SO(2) rotation around the axis in the direction of 
$\hat{x}$, i.e. the remaining symmetry after the spontaneous symmetry breaking due to the hedgehog-type VEV of the adjoint Higgs field.  
Since the generator contains not only $E_{i}$ but also $B_{i}$ through the $\theta$ term, the electric charge $q$ can be related 
to $\theta$, by imposing a condition that the theory is invariant under the aforementioned SO(2) rotation by an angle $2\pi$ (the Witten effect):
\be 
\label{4.2} 
q = ng_{4} + \frac{g_{4}\theta}{2\pi} \ \ \ \ \ (n: \ {\rm integer}).    
\ee 
So $q$ is determined by $\theta$, up to $ng_{4}$. The presence of the term $ng_{4}$ comes from the fact that 
SO(2) rotation by an angle $2\pi n$ is identity transformation.   

In the argument, however, the origin of the $\theta$ term was not specified and a possibility has been pointed out where something like the $\theta$ term is induced at quantum level by the coupling of the gauge field to fermions.

In this paper, we propose a simple model to realize the possibility in the framework of 5D gauge theory. We just add an SU(2) doublet fermion $\psi$ with bulk mass $M$ to the original SU(2) 5D pure gauge theory. In addition, we introduce an extra U(1) gauge field $C_{M} \ (M = \mu, y \ (\mu = 0 -3))$ and assume that $\psi$ carries a U(1) charge $e$. 
The relevant lagrangian for $\psi$ reads as 
\be 
\label{4.3} 
{\cal L}_{\psi} = \bar{\psi} \left[\gamma^{M}\left( i\partial_{M} + gA_{M} + e C_{M} \right) - M \right] \psi. 
\ee 

We discuss that the CS term with an operator 
$C_{y}F_{\mu \nu}\tilde{F}^{\mu \nu}$ is radiatively induced through loop diagram due to $\psi$ exchange \cite{ALM}, and replacing $C_{y}$  by its VEV $\langle C_y \rangle$, the $\theta$ term arises effectively, notably having quantized value of the $\theta$ parameter, again as the specific feature of the higher dimensional gauge theory.
  
One may wonder how the P and CP violating $\theta$ term can be realized in the 5D gauge theory compactified on a circle, where, e.g.,  
P symmetry is naturally expected to be preserved. 
An argument is possible that the interplay between the bulk mass $M$ of the fermion and the VEV $\langle C_y \rangle$ leads to P violation 
(see \cite{ALM} for the details).  
To be concrete, the calculated 5D ``CS term" due to the exchange of $\psi$ at the loop diagram is given as follows \cite{ALM}
(Interestingly, this result is also obtainable by use of Fujikawa's method \cite{Fujikawa} concerning the 4D chiral anomaly): 
\be 
\label{4.6}
{\cal L}_{CS} = - \frac{g^{2}}{32\pi^{3}R}\sum_{n_{KK}} \tan^{-1} \left( \frac{M}{\frac{n_{KK}}{R} + eC_{y}}  \right) 
{\rm Tr} \left( F_{\mu \nu}\tilde{F}^{\mu \nu} \right), 
\ee 
where the integer $n_{KK}$ denotes the KK mode. Though in \cite{ALM}, an anti-periodic boundary condition along the compactified space was adopted for $\psi$ by some technical reason, in this paper we adopt a periodic boundary condition, since the conclusion (\ref{4.7}) for sufficiently large $MR$, which plays a crucial role in our argument, is found to be the same for both choices of the boundary conditions.

Actually (\ref{4.6}) does not seem to be the CS term, since this is not linear in $C_y$ but seems to be a periodic function of $C_y$, just as in the case of the potential (\ref{2.14}). 
So, a question raised here is how to extract the genuine CS term, i.e. the term linear in $C_y$, from (\ref{4.6}).  
As was discussed in Section 1, the origin of the periodicity is attributed to the property that physical quantities are expected to be the functions of Wilson-loop, as a gauge invariant global operator, which has a physical meaning since the compactified circle has the non-trivial topological property as a non-simply-connected space. On the other hand, the genuine CS term, though it clearly does not have the periodicity, is gauge invariant, in the sense that its gauge transformation is written as a surface term. 
So, a natural guess is that the genuine CS term can be extracted by taking de-compactification limit $R \to \infty$, since in this limit the space-time reduces to 5D Minkowski space, where the non-trivial topological property and therefore the Wilson-loop lose their meanings, while the genuine CS term is still meaningful. For instance, the potential (\ref{2.14}) is easily known to vanish in the de-compactification limit.

%Actually (\ref{4.6}) is not exactly the CS term, since this is not linear in $C_y$ but is a periodic function of $C_y$ 
%just as in the case of the potential (\ref{2.14}) (though in the limit of $M \to \infty$, it turns out to reduce to the ordinary CS term, as we will see below). This is again the reflection of the fact that the extra-space component of the gauge field, $C_y$, may be regarded as a sort of AB phase, a feature never shared by ordinary 4D gauge theories.

%Since $\psi$ was introduced just for the purpose to induce the CS term, it would be desirable to make its effect in the low-energy theory as little as possible. From such a point of view, and also in order to extract the essence of our argument, let us take the limit $M \to \infty$ (more precisely $MR \gg 1$). It may be worth noting that a similar manipulation has been made in the model of Kim-Shifman-Vainshtein-Zakharov invisible axion \cite{Kim}. Naively speaking, in this limit the heavy fermion $\psi$ is expected to decouple from the low-energy theory. 

However, as a matter of fact, taking de-compactification limit is not necessary and sufficiently large $MR$ is quite enough to extract the CS term. 
In fact, we can confirm that when (\ref{4.6}) is Taylor-expanded in terms of $C_y$, all terms except the term linear in $C_y$ are negligible for sufficiently large $MR$, leaving only the linear term, which just takes the form of the genuine CS term:    
\be 
\label{4.7} 
{\cal L}_{CS} = \frac{eg^{2}R}{16\pi} \ C_{y} {\rm Tr} \left( F_{\mu \nu}\tilde{F}^{\mu \nu} \right).  
\ee 
Then replacing $C_y$ by $\langle C_y \rangle$ we finally obtain the effective $\theta$ term (\ref{4.1}) with 
\be 
\label{4.8} 
\theta = \pi eR \langle C_y \rangle.
\ee 

On the other hand, the 5D potential for $C_y$ is induced at the quantum level by the exchange of $\psi$ at the relevant loop diagram \cite{HIL}:   
\be 
\label{4.9} 
V(C_y) = - \frac{1}{4\pi^{2}}\frac{1}{L^{5}} \int_{0}^{\infty} 
\rmd s \ s^{3} \ \ln \left[ 1 - 2 \cos \left(eC_{y}L \right) e^{-\sqrt{s^{2}+ (ML)^{2}}} 
+  e^{-2\sqrt{s^{2}+ (ML)^{2}}} \right], 
\ee
where $L \equiv 2\pi R$.
For sufficiently large $MR$, the potential is well-approximated by 
\be 
\label{4.10}
V(C_y) \simeq \frac{M^{2}}{\pi^{2}L^{3}} e^{- ML} \cos \left(eC_{y}L \right), 
\ee 
which leads to 
\be 
\label{4.11} 
e \langle C_y \rangle = \frac{\pi}{L} = \frac{1}{2R}.
\ee 
Then, substituting (\ref{4.11}) in (\ref{4.8}), we get  
\be 
\label{4.12}
\theta = \frac{\pi}{2}.
\ee 
It is interesting to note that this result does not depend on $R$ nor $M$.  

Thus, from (\ref{4.2}) we finally reach to the conclusion: not only the magnetic charge $g_{m}$ but also the electric charge $q$ of the dyon is quantized and takes only discrete values:    
\be 
\label{4.13}
q = \left(n + \frac{1}{4}\right)g_{4}.
\ee
In this way, we have obtained the quantized and fractional electric charge $q$ of the dyon (in the unit of $g_{4}$), again as the inevitable consequence of the topological nature specific to the higher dimensional gauge theory. 

Accordingly, now the mixing angle $\mu$ is also quantized: (\ref{3.13}) and (\ref{4.13}) implies
\be 
\label{4.14} 
\tan \mu = - \left(n + \frac{1}{4}\right)\alpha \ \ \ \ \ (\alpha \equiv \frac{g_{4}^{2}}{4\pi}).   
\ee
An important result is that the dyon mass is quantized as well. Namely, by use of (\ref{2.9}) and (\ref{3.14}) the mass of the BPS dyon is expressed just by the radius of the compactified circle $R$ and the 4D gauge coupling $g_4$,   
\be 
\label{4.15}
M_{{\rm BPS}} = \frac{1}{\cos \mu} M_{{\rm TPM}} = \frac{\sqrt{1+(n+\frac{1}{4})^{2}\alpha^{2}}}{\alpha}\cdot \frac{1}{R} \ \ \ \ \ ( n: \ {\rm integer}).    
\ee

\section{The analysis of dyon solution based on gradient flow equations}

We have obtained the analytic BPS dyon solution
\bea 
A_y &=& F(\rho) v \hat{x}_{a}\frac{\tau_{a}}{2}, 
\nonumber \\ 
gA_{i} &=& G(\rho) \frac{1}{r} \epsilon_{aij} \hat{x}_{j}\frac{\tau_{a}}{2}, 
\nonumber \\ 
A_{0} &=&  J(\rho) v \hat{x}_{a}\frac{\tau_{a}}{2}, 
\label{5.1}  
\eea 
with (see (\ref{3.17}), (\ref{3.18}), (\ref{3.22})),    
\bea 
F(\rho) &=&  \coth \rho - \frac{1}{\rho}, 
\nonumber \\ 
G(\rho) &=& 1 - \frac{\rho}{\sinh \rho},  
\nonumber \\  
J(\rho) &=& \sin \mu \ F(\rho) = \sin \mu \ \left( \coth \rho - \frac{1}{\rho} \right) \ \ \ (\rho \equiv gvr \cos \mu).   
\label{5.2}  
\eea 
Unfortunately, however, the analytic solution is not known for the cases where the potential $V(A_y)$ is non-negligible, i.e. 
in the case of ``non-BPS dyon".

Thus some numerical analysis is necessitated in such general situations. 
One possible way is to solve the equations of motion for the gauge fields $F, G, J$ numerically, as was done in the original paper by Julia and Zee \cite{Julia-Zee}. A non-trivial thing in this method is that we have to choose the ``initial condition" 
at $\rho = 0$ for the functions $F, G, J$ carefully enough, so that the resultant values of these functions at $\rho = \infty$ 
(or practically, some sufficiently large $\rho_{\rm max}$), obtained by solving the equations of motion numerically, just coincide with the boundary conditions for these functions at $\rho = \infty$.

Here we adopt another tool, namely the gradient flow equations: 
\be 
\label{5.4} 
\partial_{s} \Phi = - \frac{\delta H}{\delta \Phi},    
\ee 
where $H$ is the Hamiltonian of the theory and $\Phi$ stands for a generic function to describe gauge field configuration, which is supposed to be not only $\rho$ dependent but also depends on the newly introduced variable $s$: $\Phi = \Phi (s, \rho)$. 
The essential point in this method is that the energy of the system decreases as $s$ increases and for $s \ \to \ \infty$, the fields approaches to the configurations satisfying $\frac{\delta H}{\delta \Phi} = 0$, which realize the minimum energy. In this 
method, the boundary conditions at both ends, $\rho = 0$ and $\rho = \infty$, are always kept during the $s-$evolution.     

For instance, in the case of BPS monopole, corresponding to $\mu = 0$ and therefore $J = 0$ (see (\ref{5.2})), 
$\Phi = F, \ G$ and the Hamiltonian is given in terms of $F, \ G$ as, 
\be 
H = \frac{2\pi v_{4}}{g_{4}} \int_{0}^{\infty} \ 
 \left[ \frac{G^{2}(2-G)^{2}}{\rho^{2}}+ \rho^{2} F'^{2} + 2G'^{2}+2F^{2}(1-G)^{2}
 \right] \ \rmd\rho,     
\label{5.5} 
\ee 
where $F' \equiv \frac{\rmd F(\rho)}{\rmd  \rho}, \ G' \equiv \frac{\rmd G(\rho)}{\rmd  \rho}$. 
Let us note that in this case, because of the subsidiary constraint (\ref{2.5}), i.e. $E_{i} = 0$, the lagrangian $L$ is simply related to $H$ as $H = - L$, and the minimization of $H$, $\frac{\delta H}{\delta \Phi} = 0$ just provides the solution to the equations of motion, $\frac{\delta L}{\delta \Phi} = 0$.

In fact, as is seen in Fig.{\ref{fig-BPS_monopole}, we have confirmed that the solutions for the gradient flow equations with respect to $F, \ G$ correctly reproduce the functions shown in (\ref{5.2}) and also the correct value of the monopole mass in the limit of sufficiently large $s$, with the boundary conditions, $F(0) = G(0) = 0$ and $F(\infty) = G(\infty) = 1$, being imposed.

\begin{figure}[htb]
\begin{center}
  \includegraphics[scale=0.9]{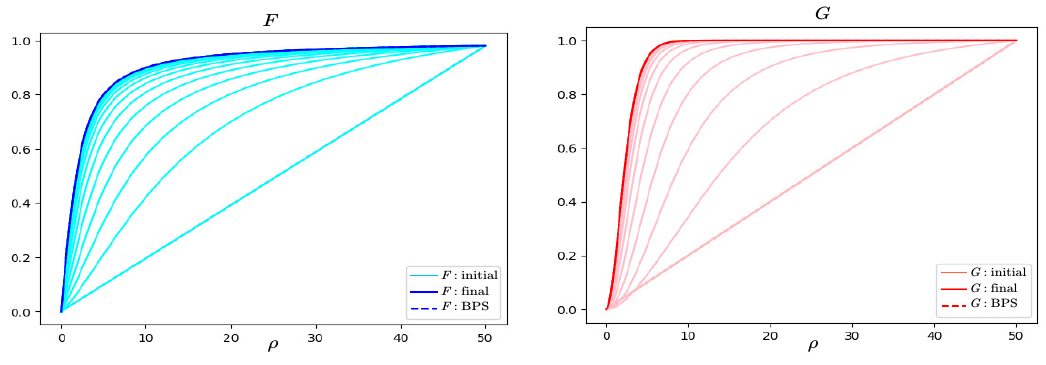}    
\end{center}
\caption{
  $F, G$ for the case of the BPS monopole. The bold solid lines denote our results, obtained starting from the initial functions, $F$: initial, etc. The dotted lines, $F$: BPS, etc., stand for the analytic BPS monopole solution.}
\label{fig-BPS_monopole}
\end{figure}

In the case of the dyon, however, the situation is a little complicated, since
in this case $E_{i}$ should exist inevitably and $H = - L$ no longer holds, even
though the fields are all static. Namely, ${\cal H} = {\rm Tr} \
\left[E_{i}^{2} + B_{i}^{2} + F_{iy}^{2} + V(A_{y})\right]$ and ${\cal L} = {\rm
Tr} \ \left[E_{i}^{2} - B_{i}^{2} - F_{iy}^{2} - V(A_{y})\right]$ leads to a relation 
\be 
\label{5.5a}
{\cal H} = 2 \ {\rm Tr} \ E_{i}^{2} - {\cal L}.  
\ee 
So, the minimization of $H$ does not yield the solutions to the equations of motion in the case of dyon. In fact, we have confirmed by direct calculation that the field configurations of BPS dyon given in (\ref{5.2}) do not satisfy $\frac{\delta H}{\delta \Phi} = 0 \ (\Phi = F, G, J)$ for vanishing $V(A_{y})$, except for the case of $\mu = 0$, corresponding to the BPS monopole. 

On the other hand, it may be worth noting that, under the Gauss law constraint $D_{i}E_{i} = 0$ (Eq.(\ref{3.2})), which is nothing but one of the equations of motion, the difference between $H$ and $- L$ turns out to be a surface term, and becomes a constant once the boundary conditions of the fields are fixed.  
To show that, first let us note that, since $E_{i} = - D_{i} A_{0}$,  
\be 
\label{5.6} 
\int \rmd^{3}x \ {\rm Tr} \ E_{i}^{2} = \int \rmd^{3}x \ {\rm Tr} \
\left(D_{i}A_{0}\right)^{2} 
= \int \rmd^{3}x \ {\rm Tr} \ \left(A_{0}D_{i}E_{i}\right) = 0,   
\ee 
where we have performed partial integral at the second equality (simply assuming that the surface term vanishes) and also used the constraint. 
This implies that the term ${\rm Tr} \ E_{i}^{2}$, if it does not vanish, should be written as a total derivative under the Gauss law constraint.

In fact, utilizing the constraint, which is explicitly written in terms of $J, \ G$ as   
\be 
\label{5.7} 
D_{i}E_{i} = 0 \ \ (\frac{\delta L}{\delta J} = 0): \ \ \ - (\rho^{2}J')'+2(1-G)^{2}J = 0,  
\ee 
we easily find that ${\rm Tr} \ E_{i}^{2}$ yields a surface term as we expected:  
\bea  
\int \ {\rm Tr} \ E_{i}^{2} \ \rmd^{3}x \rmd y &=&  \frac{1}{\cos \mu} \frac{2\pi v_{4}}{g_{4}} \int_{0}^{\infty} \ 
\rho^{2} \left[ J'^{2} + 2 \frac{(1-G)^{2}J^{2}}{\rho^{2}} \right] \rmd\rho \nonumber \\ 
&=& \frac{1}{\cos \mu} \frac{2\pi v_{4}}{g_{4}} \int_{0}^{\infty} \ 
\left[\rho^{2}J'J\right]' \ \rmd\rho = \frac{1}{\cos \mu} \frac{2\pi v_{4}}{g_{4}} \lim_{\rho \to \infty}(\rho^{2}J'J),   
\label{5.8} 
\eea 
where (\ref{5.7}) has been used to derive the second line. 

Here the boundary condition for $J$ at the spatial infinity is known to be fixed once we fix the electric charge $q$.
In fact, calculating $q$ according to (\ref{3.9}), the electric charge is given by 
\be 
\label{5.9} 
q = \frac{4\pi}{g_{4}}\frac{1}{\cos \mu} \lim_{\rho \to \infty}(\rho^{2}FJ') = g_{m} \frac{1}{\cos \mu} \lim_{\rho \to \infty}(\rho^{2}FJ'). 
\ee 
So, from the definition of the angle $\mu$ by $\tan \mu = \frac{q}{g_{m}}$, we get the boundary condition (noting $F(\infty) = 1$), 
\be 
\label{5.11}  
\lim_{\rho \to \infty}(\rho^{2}J') = \sin \mu. 
\ee 
We also get $J(\infty) = \sin \mu$ as in the case of BPS dyon, since for $\rho \to \infty$ the potential vanishes and 
the asymptotic behavior of the fields should be the same as in the case of the BPS dyon, 
\be 
\label{5.12}
{\rm for} \ \rho \to \infty, \ \ \ J(\rho) \sim \sin \mu \left[1 - \frac{1}{\rho}\right], 
\ee 
as is seen from (\ref{3.22}).

Thus we get 
\be 
\label{5.13} 
\int \ {\rm Tr} \ E_{i}^{2} \ \rmd^{3}x \rmd y = \frac{\sin^{2}\mu}{\cos \mu} \frac{2\pi v_{4}}{g_{4}},  
\ee 
and therefore a simple relation, under the constraint: 
\be 
\label{5.14}  
H = \frac{\sin^{2}\mu}{\cos \mu} \frac{4\pi v_{4}}{g_{4}} + (- L).  
\ee 
Hence, if we impose the constraint (\ref{5.7}), relating $J$ with $G$, and regard $H$ and $- L$ as functionals of two independent functions, e.g. $G$ and $F$, the minimizations of $H$ and $- L$ should result in the same field configurations, since the surface term does not contribute to the functional derivatives. 

However, when $H$ and $- L$ are regarded as the functionals of three independent functions, $F, \ G$ and $J$, we should be careful, since the extremum conditions $\frac{\delta H}{\delta \Phi}= 0$ and $\frac{\delta (- L)}{\delta \Phi}= 0$ for $\Phi = F, G, J$ yield different results. 
To be more concrete, $\frac{\delta H}{\delta \Phi}= 0$ does not yield the solutions to the equations of motion, as we have already pointed out.

Since we are interested in the solutions to the equations of motion, our task seems to solve the gradient flow equations with respect to the three independent functions, $F, G$ and $J$, where $- L$ takes place of $H$. 
By the way, only in solving the gradient flow equation, by some technical reason in the numerical computation, we adopt $H(\rho), I(\rho)$ instead of $F(\rho), J(\rho)$, defined by 
\be 
\label{5.15} 
F(\rho) \equiv \frac{H(\rho)}{\rho}, \ \ J(\rho) \equiv \frac{I(\rho)}{\rho}, 
\ee 
while keeping $G$ unchanged, although the final results are given in the form of $F, G$ and $J$. 

Then, the gradient flow equations $\partial_{s} \Phi = - \frac{\delta (-L)}{\delta \Phi}$ for $\Phi = I, G, H$ read as 
\bea 
\partial_{s} I 
&=& - C \left[I'' -2\frac{I}{\rho^2}(1-G)^2\right], 
\label{5.20a} \\  
\partial_{s} G  
&=& - 2C \left[\cos^2\mu\left\{-G''+\frac{G(2-G)(1-G)}{\rho^2}\right\}-(1-G)\frac{-I^2+H^2}{\rho^2} \right],  
\label{5.20b} \\  
\partial_{s} H  
&=& - C \left[-H'' +2\frac{H}{\rho^2}(1-G)^2 + \frac{1}{g_{4}^2v_{4}^4\cos^2\mu}\rho^2\frac{d V_4}{dH}\right],  
\label{5.20c} 
\eea 
with  
\bea 
-L &=& \frac{C}{2} \int_0^\infty{\rm d}\rho \Bigg[ -\left(I'-\frac{I}{\rho}\right)^2-2\frac{I^2}{\rho^2}(1-G)^2 +\cos^2\mu \left(2G'^2+\frac{G^2(2-G)^2}{\rho^2}\right) \nonumber \\ 
 ~~ && +\left(H'-\frac{H}{\rho}\right)^2+2\frac{H^2}{\rho^2}(1-G)^2 +\frac{2\rho^2}{g_{4}^2v_{4}^4\cos^2\mu }V_{4}(H/\rho) \Bigg], 
\label{5.21} 
\eea 
where the coefficient $C \equiv \frac{1}{\cos \mu} \frac{4\pi v_{4}}{g_{4}}$ is nothing but $M_{{\rm BPS}}$, but 
in principle it may take other positive values as well.    
As the Higgs potential, we adopt (\ref{2.14}), multiplied by $L = 2\pi R$ to get the potential in our 4D space-time:
\be 
\label{5.21a} 
V_{4}(F) = V_{4}(H/\rho) =  \frac{3g_{4}^4v_{4}^4}{8\pi^6} \sum_{w=1}^\infty\frac{1}{w^5}\left[1+2\pi RMw +\frac{1}{3}
 \left(2\pi RMw \right)^2\right] e^{-2\pi RMw} \left[ \cos \left(\pi w F\right) - (-1)^{w} \right],    
\ee
where $gv =  g_{4}v_{4} = \frac{1}{R}$ has been partially used, and $A^{(3)}_y$ in (\ref{2.14}) has been replaced by 
$|A_y| = \sqrt{\left(A^{(a)}_y\right)^{2}} =  \sqrt{(vF\hat{x}^{a})^{2}} = v F$.

In solving the gradient flow equations, the following boundary conditions are imposed: 
\bea 
&& {\rm at} \ \rho = 0: \ \ I(0) = G(0) = H(0) = 0,  
\label{5.22a} \\
&& {\rm at} \ \rho = \rho_{{\rm max}}: \ \ I'(\rho_{{\rm max}}) = \sin \mu, \ G(\rho_{{\rm max}}) = H'(\rho_{{\rm max}}) = 1,  
\label{5.22b} 
\eea 
for the numerical computation in the range of $0 \leq \rho \leq \rho_{{\rm max}}$.  

Unfortunately, by solving the gradient flow equations (\ref{5.20a}) to (\ref{5.20c}) numerically , we have found that the function $I$ shows divergent behavior as $s$ increases. The origin of this problem seems to come from the fact that in the $-L$ the term $- E_{i}^{2}$ has a negative metric, which, when  $-L$ is written in the quadratic form of the small deviations $\delta \Phi \ (\Phi = I, G, H)$ from the stationary point $\Phi_{st}$ (for which the r.h.s.s of (\ref{5.20a}) to (\ref{5.20c}) vanish) as $\Phi = \Phi_{st}+\delta \Phi$, probably leads to a negative eigenvalue of the matrix to describe the quadratic form and therefore the instability of the stationary point.  

In order to overcome this problem, since the problematic term $- E_{i}^{2}$ is mainly described by $I$, we think about the possibility to evade  
the equation (\ref{5.20a}) and replace it by imposing the analytic Gauss law constraint (\ref{5.7}) , now written in terms of $I$ as 
\be 
\label{5.22c}
I'' -2\frac{I}{\rho^2}(1-G)^2 = 0, 
\ee 
for the purpose to express $I$ in terms of $G$ and substitute such obtained $I$ in the r.h.s. of the remaining gradient flow equation (\ref{5.20b}). Nice thing of this method is once $I$ is enforced to be related to $G$ by use of the constraint, $-L$ should not have the problem of negative eigenvalue mentioned above. 
This is because, under the Gauss law constraint, $-L$ differs from $H$ just by a constant (as is seen in (\ref{5.14})), while $H$ is clearly positive definite and does not possess the problem of negative eigenvalue.

However, once again we encounter another problem. Namely, in order to express $I$ in terms of $G$ by using the constraint (\ref{5.22c}), we have to solve the differential equation of $I$ with respect to $\rho$, though it may not be impossible. 
Hence, we take another approach. Namely, in order to get $I$ for given $G$, we introduce the following another type of gradient flow equation by introducing another variable $s'$: 
\be 
\label{5.23} 
\partial_{s'} I 
= C \left[I'' -2\frac{I}{\rho^2}(1-G)^2\right],
\ee 
where $I$ is regarded to depend on the newly introduced variable $s'$ in addition to $s, \rho$: $I(s',s, \rho)$ and 
$G(s, \rho)$ is treated as a given function independent of $s'$. Then, solving this equation, $I(\infty, s, \rho)$ should satisfy the constraint (\ref{5.22c}), which we identify with $I(s, \rho)$: $I(s, \rho) =I(\infty, s, \rho)$. Then we substitute such obtained $I(s, \rho)$ into the r.h.s. of (\ref{5.20b}), and solve the remaining gradient flow equations (\ref{5.20b}) and (\ref{5.20c}) to determine the two independent functions, $G$ and $H$. 
One remark is that in (\ref{5.23}) the overall sign of the r.h.s. is just opposite to that in (\ref{5.20a}). 
Therefore the aforementioned divergence problem should be evaded. 

As we will see below, this method of modified ``two-step gradient flow equations" works well, and we obtain expected behavior of the field configurations, $F, G, J$, correctly. 
Once these field configurations are obtained, we also can calculate the mass of non-BPS dyons, say $M_{{\rm dyon}}$, numerically. For that purpose, we utilize a useful relation, which is derived from the manipulation made in (\ref{3.5}) and referring (\ref{3.10}) and (\ref{3.14}):  
\be 
\label{5.24}
M_{{\rm dyon}} = \frac{1}{\cos \mu} \frac{4\pi v_{4}}{g_{4}} + \tilde{H} =  M_{{\rm BPS}}  + \tilde{H}, 
\ee  
where 
\bea 
\tilde{H} &=& \int \ \Big[{\rm Tr} \{ E'^{2}_{i} + (B'_{i} + F_{iy})^{2} \} +
V(H/\rho) \Big] \ \rmd^{3}x \rmd y \nonumber \\ 
&=& \frac{M_{{\rm BPS}}}{2} \int_0^\infty \rmd\rho \Bigg[ 2\cos^2\mu \left\{\sin\mu G'-\frac{I}{\rho}(1-G)\right\}^2 +\cos^2\mu \left\{I'-\frac{I}{\rho} +\sin\mu \frac{G(G-2)}{\rho}\right\}^2 \nonumber \\ 
&&+ 2\left\{\cos^2\mu G'-\frac{H-\sin\mu I}{\rho}(1-G)\right\}^2 \nonumber \\ 
&&+ \left\{H'-\sin\mu I'-\frac{H-\sin\mu I}{\rho}+\cos^2\mu \frac{G(G-2)}{\rho}\right\}^2 +\frac{2\rho^2}{g_{4}^2v_{4}^4\cos^2\mu}V_{4}(H/\rho) \Bigg],   
\label{5.25} 
\eea 
which is clearly positive-semidefinite and exactly vanishes for the BPS dyon. Of course, we may also calculate $M_{{\rm dyon}}$ 
by directly integrating the Hamiltonian density. In the direct computation, however, the piece corresponding to $M_{{\rm BPS}}$, which is the linear combination of $g_m$ and $q$ and therefore behaves as a surface term, should have an error due to the limitation of the integration range by the finiteness of $\rho_{{\rm max}}$. One of the merit of our procedure is that 
the piece is replaced by its theoretical exact value without the error.

We are now ready to show some of our results, obtained by solving the modified two-step gradient flow equations. 
First, in order to demonstrate the validity of the method, we calculated $F, G$ and $J$ for the case of BPS dyon with 
$\mu = 1$, for instance, whose results are shown in Fig.\ref{fig-BPS_dyon}. In this computation, $\rho_{{\rm max}} = 20$ is taken.

\begin{figure}[htb]
\begin{center}
  \includegraphics[scale=0.8]{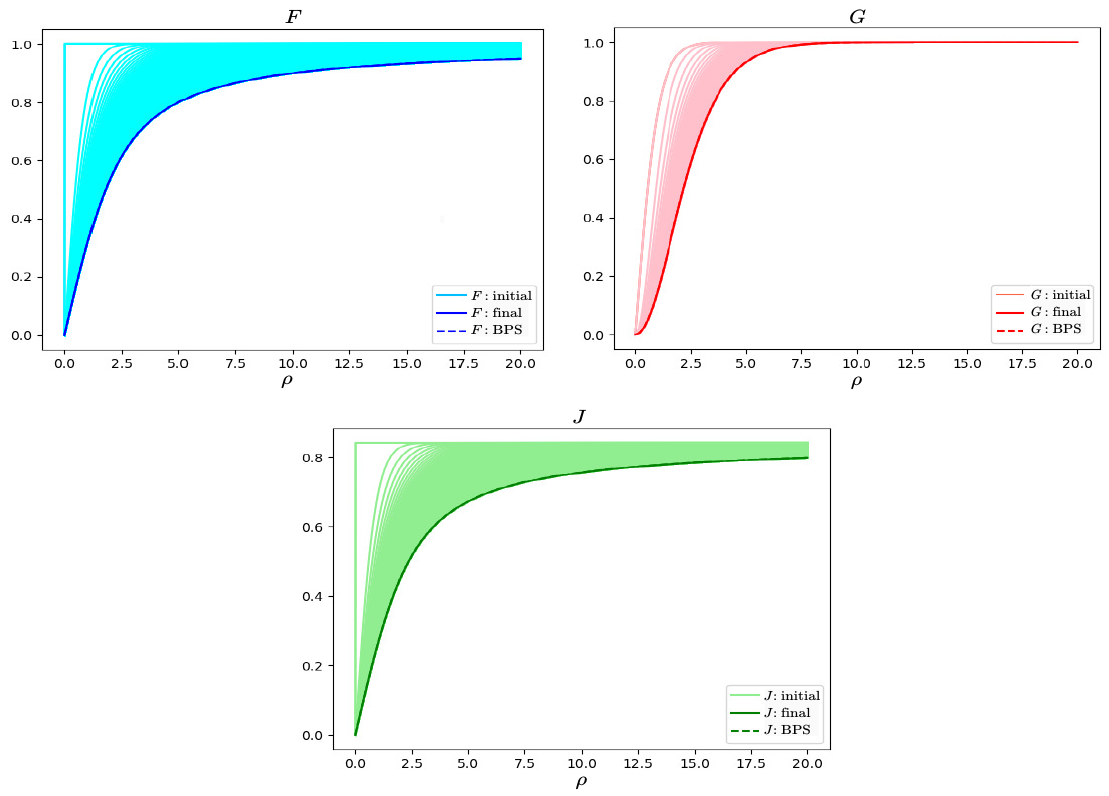}    
\end{center}
\caption{
  $F, G, J$ for the case of the BPS dyon with $\mu = 1$.
  The bold solid lines denote our results, obtained starting from the initial functions, $F$: initial, etc. The dotted lines, $F$: BPS, etc., stand for the BPS dyon solution.}
\label{fig-BPS_dyon}
\end{figure} 

As is seen in these figures, the coincidence between the calculated field configurations $F, G$ and $J$ and their theoretical curves is just perfect, and actually we cannot distinguish them in the Figure. 
Also, calculated $M_{{\rm dyon}}$ is in quite good agreement with $M_{{\rm BPS}}$. 

Next we consider the realistic non-BPS dyons with the GHU potential (\ref{5.21a}) being switched on. 
For illustration, we take only two cases: (1) $MR = 1$ and (2) $M = 0$. Concerning the parameter $\mu$, we adopt 
the prediction of our higher dimensional theory, $\tan \mu = - \frac{g_{4}^{2}}{16\pi}$ (corresponding to $n = 0$ in (\ref{4.14})), though for brevity we set $g_{4} = 1$. In all computational results shown hereafter we have taken $\rho_{{\rm max}} = 50$.

First we consider the case (1). An important observation here is that for sufficiently large bulk mass, or more precisely 
sufficiently large $MR$, as was assumed in the section 4 above, the GHU potential gets exponential suppression: 
$V_{4}(F) \simeq \frac{g_{4}^{4}v_{4}^{4}}{2\pi^{4}} (RM)^{2} e^{-2\pi RM}[\cos (\pi F) + 1]$. 
Thus, even in this case, where $MR$ is not so large, we anticipate that $F, G, J$ are close to the theoretical curves of the BPS monopole. The results for the field configurations are shown in Fig.\ref{fig-nonBPS_MR=1}.  

\begin{figure}[htb]
\begin{center}
  \includegraphics[scale=0.8]{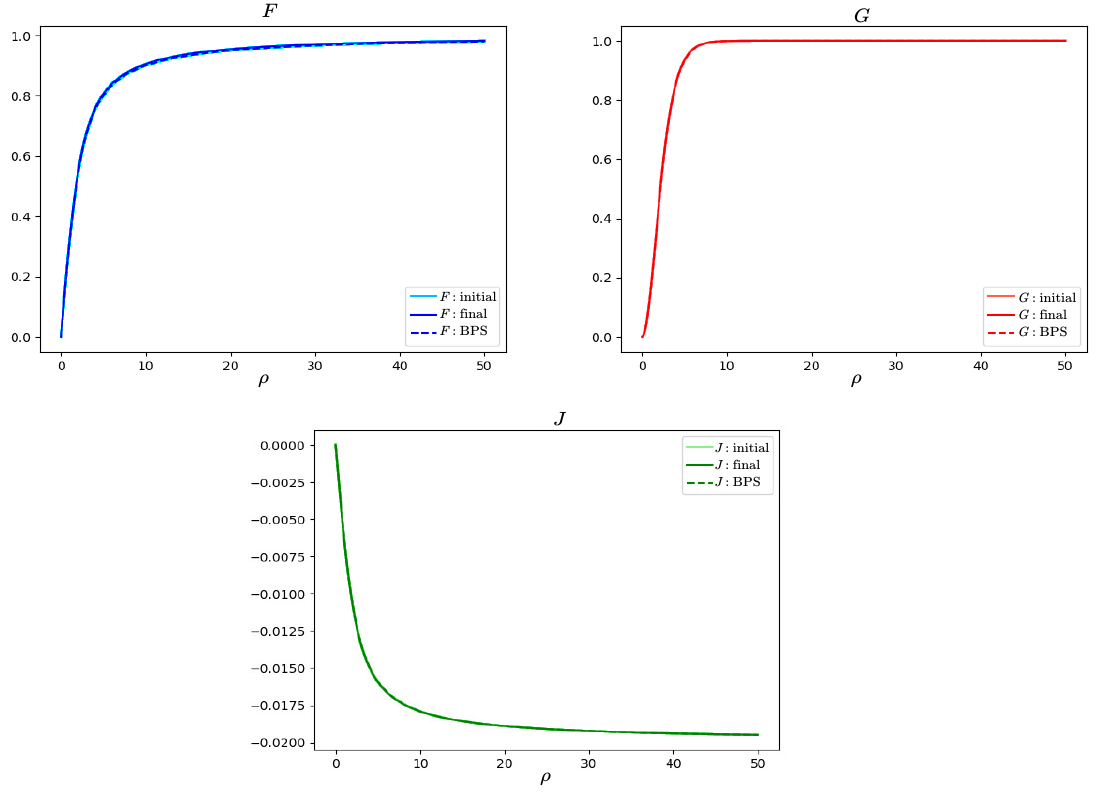}    
\end{center}
\caption{$F, G, J$ for the case of the non-BPS dyon with $MR = 1$.The bold solid lines denote our results, obtained starting from the initial functions, $F$: initial, etc. The dotted lines, $F$: BPS, etc., stand for the BPS dyon solution.}
\label{fig-nonBPS_MR=1}
\end{figure} 

As is seen in these figures, the deviations of the functional forms of $F, G, J$ from the case of BPS dyon are very small, and, in particular, concerning $G$ and $J$ the differences are practically unrecognizable. 
Accordingly, the calculated $M_{{\rm dyon}} = 1.003 M_{{\rm BPS}}$ is only slightly larger than $M_{{\rm BPS}}$, reflecting the small deviations of the functions. 

Next we consider the case (2). In this case, there is no exponential suppression factor: $V_{4}(F) = 
\frac{3g_{4}^4v_{4}^4}{8\pi^6} \sum_{w=1}^\infty\frac{1}{w^5} \left[ \cos \left(\pi w F\right) - (-1)^{w} \right]$. 
The results for the field configurations are shown in Fig.\ref{fig-nonBPS_M=0}. 

\begin{figure}[htb]
\begin{center}
  \includegraphics[scale=0.75]{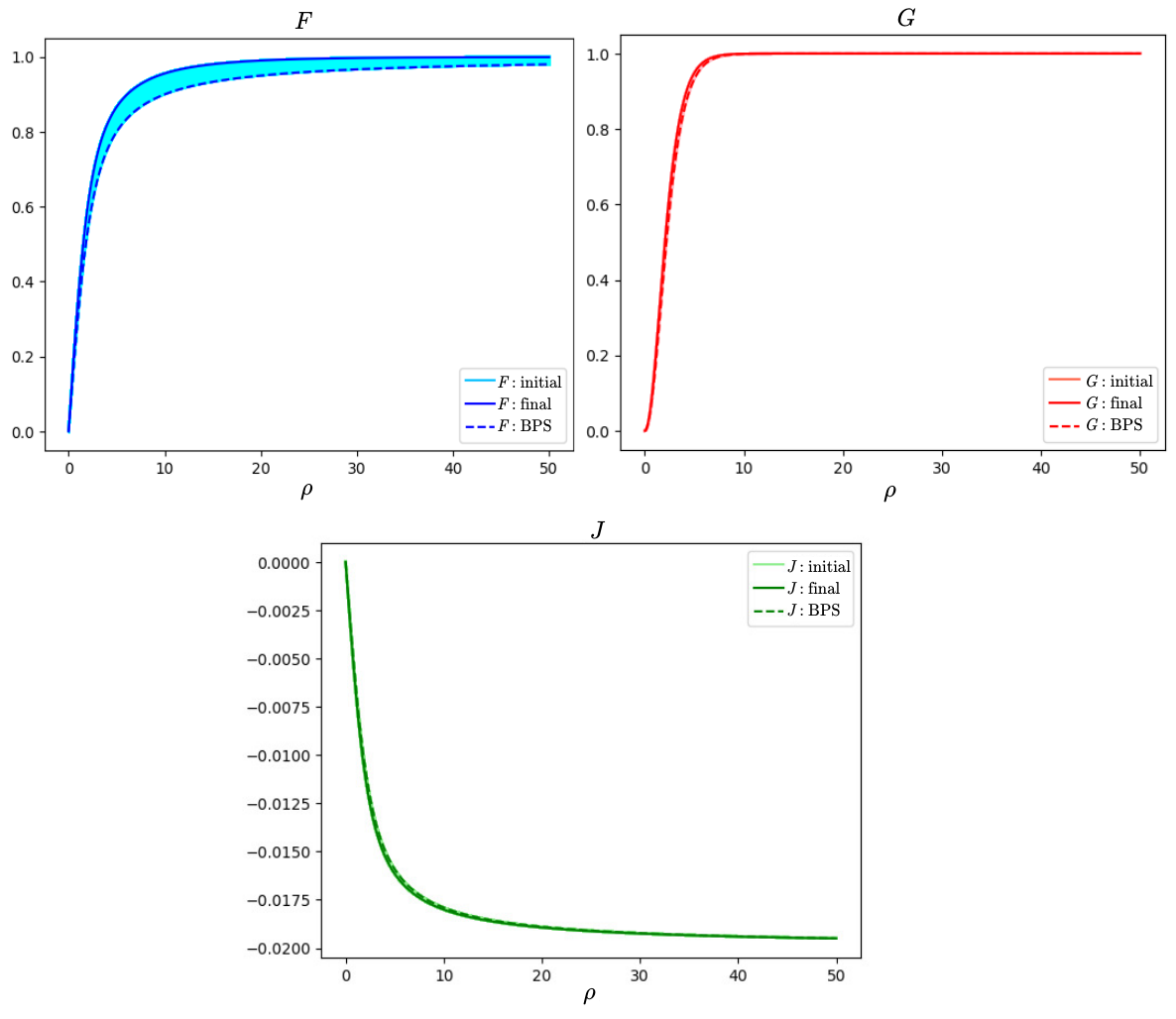}    
\end{center}
\caption{$F, G, J$ for the case of the non-BPS dyon with $M = 0$.The bold solid lines denote our results, obtained starting from the initial functions, $F$: initial, etc. The dotted lines, $F$: BPS, etc., stand for the BPS dyon solution.}
\label{fig-nonBPS_M=0}
\end{figure}

As is seen in these figures, the functional form of $F$ deviates from the case of BPS dyon, though the deviation is not large. Concerning $G$ and $J$, the functional forms still almost coincide with those predicted for the BPS dyon. 
The calculated $M_{{\rm dyon}} = 1.019 M_{{\rm BPS}}$ is still not so different from $M_{{\rm BPS}}$.

Although we have shown the results only for two typical cases, it will be safe to state that the deviations of the functional forms of $F, G, J$ and the mass $M_{{\rm dyon}}$ from what we expect in the case of the BPS dyon are relatively small and the analytically obtained BPS solution and its mass are reasonably good approximations. This probably reflects the fact that the Higgs potential is radiatively induced at the quantum level in the GHU scenario.

\section{Summary and discussion}

In this paper, we discussed the TP monopole and then dyon in the framework of the higher dimensional gauge theories, such as the scenario of GHU. In these theories, the adjoint Higgs 
field, indispensable for the construction of TP monopole or dyon, has been automatically included in the theory as the extra space component of the higher dimensional gauge field. 
 
As a specific feature of this type of theories, especially the 5D GHU compactified on a circle, 
the Higgs potential is periodic in the Higgs field, reflecting the topological nature of the extra dimension. 
This property was argued to play central roles in deriving the multiple characteristic predictions made by the higher dimensional gauge theory, demonstrated in this paper.

First, we argued that, as a specific feature of higher dimensional gauge theory such as 5D SU(2) GHU, the BPS saturated TP monopole, say BPS monopole, is nothing but an (anti-)self-dual gauge field in the 4D space with coordinates $(\vec{x}, y)$, where $y$ denotes the coordinate of the extra space. 
Furthermore, it was demonstrated that the mass of the BPS monopole $M_{{\rm TPM}}$ is quantized and takes only discrete values, just as the action of the instanton is quantized, in clear contrast to the case of the theories in ordinary 4D space-time, where $M_{{\rm TPM}}$ is proportional to the Higgs VEV, which is left as a continuous free parameter, since the Higgs potential to fix the VEV is absent for the BPS monopole. 
In other words, this means that the Higgs VEV is also quantized in the GHU scenario, and it was pointed out that this conclusion is endorsed from a different view point, relying on the periodicity of the Higgs potential.

As was discussed in some details in section 2, from the viewpoint of the caloron \cite{Harrington, Lee, Dunne}, which may be understood to be a composition of a pair of constituent monopole and anti-monopole, our conclusion of the quantized monopole mass corresponds to the specific case, where only a single monopole exists that carries the whole topological charge.

Next, the argument was generalized to the case of dyon. It turns out that one of the BPS conditions still can be written as an (anti-)self-dual-like condition, similar to that in the case of BPS monopole, and the mass of the BPS dyon, $M_{{\rm BPS}}$ is still proportional to the quantized Higgs VEV in the higher dimensional gauge theory. 
$M_{{\rm BPS}}$, however, also depends on a parameter $\mu$, denoting the ratio of the electric and magnetic charges of 
the dyon, which is usually left as a free parameter.

In this paper, we proposed an interesting mechanism to quantize the parameter $\mu$, again utilizing the characteristic features of the higher dimensional gauge theories. Namely, we showed that in the 5D gauge theories, the CS term is induced at the quantum level, and replacing the extra space component of the introduced U(1) gauge field by its VEV, the $\theta$ term arises. 
Importantly, here again the VEV is quantized and through the Witten effect we have reached to a conclusion that the electric charge of the dyon, and therefore the parameter $\mu$ is discretized. This means that $M_{{\rm BPS}}$ is also discretized. 

We also proposed a numerical method to obtain the field configurations of the non-BPS dyons, whose analytic forms are not known. 
What we proposed was the method to utilize ``modified" gradient flow equations to get the solutions of the equations of motion. 
The method was confirmed to work well, and we have found that in the 5D gauge theory, the field configurations of non-BPS dyon and 
its mass, obtained by solving the modified gradient flow equations numerically, are rather close to those of the BPS dyon, probably reflecting the fact that the Higgs potential is induced only at the quantum level in the scenario of GHU.    

The higher dimensional gauge theories such as 5D SU(3) GHU were originally proposed in order to solve the gauge hierarchy problem in the unified electro-weak theories \cite{HIL, Kubo}. 
However, they may have some interesting implications also in the strongly coupled gauge theories.

For example, since the extra space component $A_y$ in 5D GHU behaves as a pseudo-scalar, like the axion, from the 4D point of 
view, the 5D theory provides a natural solution to the strong CP problem, where the CS term discussed in this paper 
behaves as the term of axion coupling with the SU(3) gauge field in ordinary 4D theories \cite{ALM}.

The 't Hooft anomaly \cite{'t Hooft anomaly} of ordinary and/or generalized symmetries, has been known to play a critical role 
in the investigation of the symmetries in the low-energy effective theories of strongly coupled theories, such as QCD 
\cite{Tanizaki-Kikuchi}. Here again, 5D gauge theories provide a natural framework to understand the 't Hooft anomaly, regarding the QCD as a theory on the 4D boundary of 5D CS theory, through the mechanism of anomaly inflow \cite{anomaly inflow}.  

On the other hand, the color confinement in the QCD is able to be intuitively understood as the result of the dual Meissner effect 
and the monopole condensation. In the mechanism, because of the presence of the $\theta$ term to describe the non-trivial nature of the QCD vacuum, dyon condensation is expected to play an important role \cite{Hayashi-Tanizaki}, through the Witten effect \cite{Witten}. 

It will be nice if the findings in this paper, based on the characteristic properties of higher dimensional gauge theories, 
have some implications for these issues.

\subsection*{Acknowledgments}

This work was supported in part by Japan Society for the Promotion of Science, Grants-in-Aid for Scientific Research, 
No.~25K07304(NM).  
%No.~16H00872.

%\bibliography{./GHU_ref_library} 
%\bibliographystyle{./JHEP} 

%%%%%%%%%%%%%%%%%%%%
%%%% From here
%%%%%%%%%%%%%%%%%%%%
\providecommand{\href}[2]{#2}\begingroup\raggedright\endgroup
%%%%%%%%%%%%%%%%%%%%
%%%% Till here
%%%%%%%%%%%%%%%%%%%%

\end{document}